\input lanlmac.tex
\input epsf


\def\unlockat{\catcode`\@=11}
\def\lockat{\catcode`\@=12}

\unlockat

\def\newsec#1{\global\advance\secno by1\message{(\the\secno. #1)}
\global\subsecno=0\global\subsubsecno=0\eqnres@t\noindent
{\bf\the\secno. #1}
\writetoca{{\secsym} {#1}}\par\nobreak\medskip\nobreak}
\global\newcount\subsecno \global\subsecno=0
\def\subsec#1{\global\advance\subsecno
by1\message{(\secsym\the\subsecno. #1)}
\ifnum\lastpenalty>9000\else\bigbreak\fi\global\subsubsecno=0
\noindent{\it\secsym\the\subsecno. #1}
\writetoca{\string\quad {\secsym\the\subsecno.} {#1}}
\par\nobreak\medskip\nobreak}
\global\newcount\subsubsecno \global\subsubsecno=0
\def\subsubsec#1{\global\advance\subsubsecno by1
\message{(\secsym\the\subsecno.\the\subsubsecno. #1)}
\ifnum\lastpenalty>9000\else\bigbreak\fi
\noindent\quad{\secsym\the\subsecno.\the\subsubsecno.}{#1}
\writetoca{\string\qquad{\secsym\the\subsecno.\the\subsubsecno.}{#1}}
\par\nobreak\medskip\nobreak}

\def\subsubseclab#1{\DefWarn#1\xdef
#1{\noexpand\hyperref{}{subsubsection}%
{\secsym\the\subsecno.\the\subsubsecno}%
{\secsym\the\subsecno.\the\subsubsecno}}%
\writedef{#1\leftbracket#1}\wrlabeL{#1=#1}}
\lockat

\def\IL{\relax{\rm I\kern-.18em L}}
\def\IH{\relax{\rm I\kern-.18em H}}
\def\IR{\relax{\rm I\kern-.18em R}}
\def\IC{\relax\hbox{$\inbar\kern-.3em{\rm C}$}}
\def\IZ{\relax\ifmmode\mathchoice
{\hbox{\cmss Z\kern-.4em Z}}{\hbox{\cmss Z\kern-.4em Z}}
{\lower.9pt\hbox{\cmsss Z\kern-.4em Z}}
{\lower1.2pt\hbox{\cmsss Z\kern-.4em Z}}\else{\cmss Z\kern-.4em
Z}\fi}


\font\manual=manfnt \def\dbend{\lower3.5pt\hbox{\manual\char127}}

\def\IZ{\relax\ifmmode\mathchoice
{\hbox{\cmss Z\kern-.4em Z}}{\hbox{\cmss Z\kern-.4em Z}}
{\lower.9pt\hbox{\cmsss Z\kern-.4em Z}}
{\lower1.2pt\hbox{\cmsss Z\kern-.4em Z}}\else{\cmss Z\kern-.4em
Z}\fi}


\def\IZ{\relax\ifmmode\mathchoice
{\hbox{\cmss Z\kern-.4em Z}}{\hbox{\cmss Z\kern-.4em Z}}
{\lower.9pt\hbox{\cmsss Z\kern-.4em Z}}
{\lower1.2pt\hbox{\cmsss Z\kern-.4em Z}}\else{\cmss Z\kern-.4em
Z}\fi}
\def\IB{\relax{\rm I\kern-.18em B}}
\def\IC{{\relax\hbox{$\inbar\kern-.3em{\rm C}$}}}
\def\ID{\relax{\rm I\kern-.18em D}}
\def\IE{\relax{\rm I\kern-.18em E}}
\def\IF{\relax{\rm I\kern-.18em F}}
\def\IG{\relax\hbox{$\inbar\kern-.3em{\rm G}$}}
\def\IGa{\relax\hbox{${\rm I}\kern-.18em\Gamma$}}
\def\IH{\relax{\rm I\kern-.18em H}}
\def\II{\relax{\rm I\kern-.18em I}}
\def\IK{\relax{\rm I\kern-.18em K}}
\def\IP{\relax{\rm I\kern-.18em P}}

\def\inbar{\,\vrule height1.5ex width.4pt depth0pt}

\font\cmss=cmss10 \font\cmsss=cmss10 at 7pt
\def\IR{\relax{\rm I\kern-.18em R}}


\def\boxit#1{\vbox{\hrule\hbox{\vrule\kern8pt
\vbox{\hbox{\kern8pt}\hbox{\vbox{#1}}\hbox{\kern8pt}}
\kern8pt\vrule}\hrule}}
\def\mathboxit#1{\vbox{\hrule\hbox{\vrule\kern8pt\vbox{\kern8pt
\hbox{$\displaystyle #1$}\kern8pt}\kern8pt\vrule}\hrule}}


\def\inbar{\,\vrule height1.5ex width.4pt depth0pt}

\font\cmss=cmss10 \font\cmsss=cmss10 at 7pt
\def\IR{\relax{\rm I\kern-.18em R}}

\Title{ \vbox{\baselineskip12pt\hbox{hep-th/98}
\hbox{YCTP-P24-98}
\hbox{TMUP-HEL-9813}}}
{\vbox{
\centerline{Cooper pairing at large $N$ in a $2$-dimensional model}
}}\footnote{}
\medskip
\centerline{Alan Chodos $^1$, Hisakazu Minakata $^2$ and Fred Cooper$^3$}
\medskip
\centerline{$^1$Center for Theoretical Physics, Yale University}
\centerline{P.O. Box 208120, New Haven, CT 06520-8120 USA}
\medskip
\centerline{$^2$Department of Physics, Tokyo Metropolitan University}
\centerline{1-1 Minami-Osawa, Hachioji, Tokyo 192-0397, Japan}
\medskip
\centerline{$^3$Theoretical Division, Los Alamos Scientific Laboratory}
\centerline{Los Alamos, New Mexico 87455, and}
\centerline{Physic Department, Boston College}
\centerline{Higgins Hall, 140 Commonwealth Avenue, Chestnut Hill, MA
02167-3811}

\bigskip
\centerline{{\bf Abstract}}

We study a $2$-dimensional model of fermi fields $\psi$ that is closely
related to the Gross-Neveu model, and show that to leading order in ${1
\over N}$ a $\langle \psi\psi \rangle$ condensate forms. This effect is
independent of the chemical potential, a peculiarity that we expect to be
specific to $2$ dimensions. We also expect the condensate to be unstable
against corrections  at higher orders in ${1 \over N}$. We compute the
Green's functions associated with the composite $\psi\psi$, and show that
the fermion acquires a Majorana mass proportional to the gap, and that a
massless Goldstone pole appears.

\bigskip
\bigskip
Recently several papers have appeared [1-3] dealing with the properties of
QCD at high density. The basic procedure is to approximate QCD by a direct
four-quark interaction term, justifying this either by appeal to instanton
effects or to one-gluon exchange. In the presence of a chemical potential,
this theory admits a condensate of quark-quark pairs, very similar to the
Cooper pairs that are well-known in the BCS theory of superconductivity.
This phenomenon, which has been dubbed "color superconductivity," may or
may not be accessible to experiments on heavy-ion collisions that will be
performed over the next few years.

In this paper we shall examine similar phenomena in the context of a
one-plus-one dimensional model that is a close relative to the Gross-Neveu
(GN) model [4]. From its inception, it has been recognized that the GN
model exhibits many of the same features as QCD, such as asymptotic freedom
and spontaneous chiral symmetry breaking. Moreover, the four-fermi
interaction $\underline{is}$ the model - there is no need to regard it as
an approximation to an underlying gauge theory. Unlike in higher
dimensions, in two dimensions this interaction is renormalizable, and we
shall find, just as in the usual GN model, that coupling constant
renormalization removes all the divergences that we shall encounter.
Another advantage is that by judiciously introducing a flavor index $i$, $1
\leq i \leq N$, one can insure that the mean-field approximation that is
commonly used in analyzing the condensate in QCD is in the case of the
$2$-dimensional model justified as the leading contribution in powers of $1/N$.

There are, however, a couple of peculiarities associated with two
dimensions that make this GN-like model qualitatively different from QCD.
The first is the Coleman-Mermin-Wagner theorem [5], which forbids
spontaneous symmetry breaking of a continuous symmetry. Whereas in the
original GN model the broken symmetry is a discrete one, and hence not in
conflict with the theorem, in this case the formation of a $\langle \psi
\psi \rangle$ condensate breaks fermion number, a continuous symmetry. The
same problem arises in the chiral GN model [4], where the symmetry is
continuous, and in a variety of other two-dimensional models where the
spontaneous breaking of a continuous symmetry is predicted in leading order
in $1/N$. This means that instabilities must arise in higher order that
vitiate the prediction of a condensate.  However, as Witten has pointed out
[6], the $1/N$ expansion may still be an excellent guide to the physics of
the model, except for the formation of the condensate (what happens in
these models is that the condensate "almost" forms, in the sense that the
pair-pair correlation function decays in the infrared only like a power
instead of exponentially, and the power vanishes as $N \rightarrow \infty$).

The second peculiarity is, as we shall show below, the chemical potential
has nothing to do with the formation of the condensate. In higher
dimensions the chemical potential is crucial, because it gives rise to the
Fermi surface at which the gap equation has an infrared singularity as the
gap goes to zero. It is this feature that insures that the gap equation
will have a solution for arbitrarily weak coupling. In two dimensions,
however, the Fermi surface has dimension zero, and the infrared singularity
exists whether or not there is a chemical potential. In fact, the gap
equation turns out to be completely independent of the chemical potential.
This behavior will be exhibited explicitly below.

The model we consider is defined by the following Lagrangian:

$$
{\cal L} = \bar{\psi}^{(i)} i \not{\partial} \psi^{(i)} + 2 g^2
\bar{\psi}^{(i)} \gamma_5 \psi^{(j)} \bar{\psi}^{(i)} \gamma_5 \psi^{(j)}
~. \eqno(1)
$$

\bigskip\noindent
$\psi^{(i)}$ is a two-component spinor with a flavor index that takes on
$N$ values. Repeated flavor indices in eqn. (1) are summed. Because of the
unconventional arrangement of flavor indices in the second term, the model
does not have $SU(N)$ symmetry, but it does possess $O(N)$ symmetry,
$\psi^{(i)} \rightarrow \vartheta^{ij} \psi^{(j)}$ where $\vartheta^{ij}$
is a real $N \times N$ orthogonal matrix.  The Lagrangian (1) also has a
$U(1)$ symmetry, which we shall find is broken by a $\langle \psi\psi
\rangle$ condensate, whereas the $O(N)$ symmetry is kept intact.

Our representation for the $\gamma$-matrices is: ~$\gamma^0 = \sigma_1;
~\gamma^1 = - i \sigma_2; ~\gamma_5 = \sigma_3$, and it is then easy to
check that

$$
\bar{\psi}^{(i)} \gamma_5 \psi^{(j)} \bar{\psi}^{(i)} \gamma_5 \psi^{(j)} =
- {1 \over 2} (\epsilon_{\alpha\beta} \psi_{\alpha}^{\dagger(i)}
\psi_{\beta}^{\dagger(i)}) (\epsilon_{\gamma\delta} \psi_{\gamma}^{(j)}
\psi_{\delta}^{(j)}) ~. \eqno(2)
$$

\bigskip\noindent
Following the usual Hubbard-Stratonovich procedure, in the form introduced
by Coleman [7], we add to ${\cal L}$ the term

$$
- ~{1 \over g^2} (B^{\dagger} - g^2 ~\epsilon_{\alpha\beta}
\psi_{\alpha}^{\dagger(i)} \psi_{\beta}^{\dagger(i)}) (B + g^2
~\epsilon_{\gamma\delta} \psi_{\gamma}^{(j)} \psi_{\delta}^{(j)}) \eqno(3)
$$

\bigskip\noindent
which does not affect the physics because $B$ and $B^{\dagger}$ are simply
auxiliary fields. We then have

$$
{\cal L} = \bar{\psi}^{(i)} (i \not{\partial} - \mu \gamma^0) \psi^{(i)} -
{1 \over g^2} B^{\dagger} B + B ~\epsilon_{\alpha\beta}
\psi_{\alpha}^{\dagger(i)} \psi_{\beta}^{\dagger(i)} - B^{\dagger}
~\epsilon_{\alpha\beta} \psi_{\alpha}^{(i)} \psi_{\beta}^{(i)} \eqno(4)
$$

\bigskip\noindent
where we have also introduced a chemical potential $\mu$. In anticipation
of taking the large $N$ limit, $N \rightarrow \infty$ with $\lambda = g^2N$
fixed, we rewrite the second term as $- ~{N \over \lambda} B^{\dagger}B$.

Thus the classical (or tree-level) term in $V_{eff}(B^{\dagger}B)$ is of
order $N$. As we perform a perturbation expansion, we observe that
additional factors of $N$ arise in 2 ways: ~(i) the $B-B^{\dagger}$
propagator is proportional to $1/N$; and (ii) each closed fermion loop
gives a factor of $N$ from summing on the flavor index. If we examine the
computation of higher-loop corrections to the effective potential (i.e. the
summation over all one-particle-irreducible diagrams with zero-momentum $B$
and $B^{\dagger}$ external legs) we see that the one-loop term is of order
$N$ (one fermion loop and no $B-B^{\dagger}$ propagator) whereas anything
else is of order $N^0$ or lower.  Hence the leading contribution is just
what we get from keeping the tree and one-loop graphs. Moreover, in a
path-integral approach, one first integrates out the fermions. Then,
because the exponent is proportional to $N$, one can employ the stationary
phase approximation in the integral over $B$ and $B^{\dagger}$ to evaluate
the integrand at the solution of the equations

$$
{\partial {V}\over \partial B} = {\partial {V}\over \partial B^{\dagger}} =
0 ~. \eqno(5)
$$

\bigskip\noindent
The task of integrating out the fermions is complicated slightly by the
$\psi\psi$ and $\psi^{\dagger}\psi^{\dagger}$ terms. We observe that

$$
\int D\psi e^{i(\psi, {\cal M}\psi)} = det^{1/2} {\cal M} ~, \eqno(6)
$$

\bigskip\noindent
where ${\cal M}$ is any anti-symmetric matrix. We write the fermion part of
our Lagrangian as

$$
{\cal L} = {1 \over 2} (\psi^{\dagger} A \psi - \psi A^T \psi^{\dagger}) +
\psi^{\dagger} {\cal B} \psi^{\dagger} + \psi {\cal B}^{\dagger} \psi \eqno(7)
$$

\noindent
where

$$
A = (i \partial_0 + i \sigma_3 \partial_x - \mu)_{\alpha\beta} \delta^{ij}
~; \eqno(8a)
$$
$$
A^T = (- i \partial_0 - i \sigma_3 \partial_x - \mu)_{\alpha\beta}
\delta^{ij} ~; \eqno(8b)
$$
$$
{\cal B} = B \epsilon_{\alpha\beta} \delta^{ij} = i
B(\sigma_2)_{\alpha\beta} \delta^{ij} ~; \eqno(8c)
$$
$$
{\cal B}^{\dagger} = - i B^{\dagger}(\sigma_2)_{\alpha\beta} \delta^{ij} ~.
 \eqno(8d)
$$

\bigskip\noindent
Now we perform a translation,

$$
\psi = \chi + \alpha \psi^{\dagger} \eqno(9a)
$$
$$
= \chi + \psi^{\dagger} \alpha^T \eqno(9b)
$$

\bigskip\noindent
where $\alpha = {1 \over 2} ({\cal B}^{\dagger})^{-1} A^T$; this factorizes
the $\chi$ and $\psi^{\dagger}$ path integrals into the product of two path
integrals, and after some manipulations and discarding an overall factor
that is independent of ${\cal B}$ and ${\cal B}^{\dagger}$, we obtain

$$
e^{i \Gamma_{eff}^{(1)}(B, B^{\dagger})} = det^{1/2} [{\bf 1} + 4 A^{-1}
{\cal B} (A^T)^{-1} {\cal B}^{\dagger}] \eqno(10)
$$

\bigskip\noindent
where $\Gamma_{eff}^{(1)}$ denotes the one-loop contribution to the
effective action.

Note that the matrix $A^{-1}$ is given by

$$
A^{-1}(x, y) = - \int {d^2k \over (2 \pi)^2} ~{[k_0 - \sigma_3 k_1 +
\mu]_{\alpha\beta} \delta^{ij} e^{i k \cdot (x - y)} \over (k_0 + \mu +
i\epsilon ~sgnk_0)^2 - k_1^2} \eqno(11)
$$

\bigskip\noindent
where $k \cdot x = k_0 x^0 + k_1 x^1$, and the $i ~\epsilon$ prescription
is introduced in the proper way to take account of the chemical potential.
$(A^T)^{-1}$ is the same expression but with $x$ and $y$ interchanged.

To obtain the effective potential, we take $B$ and $B^{\dagger}$ to be
constant. It is then convenient to write everything in momentum space, and
after some algebra, using $\Gamma_{eff} = - V_{eff} \int d^2x$, we obtain

$$
\eqalign{
V_{eff} = &{N B^{\dagger} B \over \lambda} + {i N \over 2} \int {d^2k \over
(2\pi)^2} \cr
&tr ln \biggr{[}1 - {4 B^{\dagger} B \over [(k_0 + \mu + i ~\epsilon~
sgnk_0) + k_1 \sigma_3]  [(k_0 - \mu + i ~\epsilon~ sgnk_0) - k_1
\sigma_3]}\biggr{]} \cr
}
\eqno(12)
$$

\bigskip\noindent
where the trace is over the spinor indices only. Setting $B^{\dagger}B =
M/4$ and $\kappa = 4\lambda$, and observing that $V_{eff}(M = 0) = 0$, we
can write

$$
{1 \over N} V_{eff} = {1 \over N} \int_0^M dM^{\prime} {dV \over
dM^{\prime}} ~, \eqno(13)
$$

\noindent
where

$${1 \over N} ~{dV \over dM} = {1 \over \kappa} + {i \over 2} \int {d^2k
\over (2\pi)^2} ~tr~ \biggr{[}{1 \over M - k_0^2 + (k_1 \sigma_3 + \mu)^2 -
i ~\epsilon}\biggr{]} ~. \eqno(14)
$$

\bigskip\noindent
Here the trace is just summation over $\sigma_3 = \pm 1$.

We note that this integral is logarithmically divergent. We shall deal with
this by renormalizing $\kappa$;  but first we shall do the $k_0$ integral.
Let $M + (k_1 \sigma_3 + \mu)^2 = \omega^2$ with $\omega > 0$. We have

$$
I = \int_{- \infty}^{\infty} dk_0 {1 \over k_0 - \omega^2 + i \epsilon} =
\int_{- \infty}^{\infty} dk_0 {1 \over (k_0 - \omega + i \epsilon)(k_0 +
\omega - i \epsilon)}
$$

\medskip
$$= - {\pi i \over \omega}  \eqno(15)
$$

\noindent
so

$$
{1 \over N} {dV \over dM} = {1 \over \kappa} - {1 \over 8 \pi} \int_{-
\infty}^{\infty} dk_1 tr ({1 \over \omega}) . \eqno(16)
$$

\bigskip\noindent
We renormalize by requiring that at $\mu = 0$,

$$
{1 \over N} ~{\partial^2 V_{eff} \over \partial B \partial B^{\dagger}}
\mid_{B^{\dagger}B = M_0/4} = {4 \over \kappa_R} ~. \eqno(17)
$$

\noindent\bigskip
which is the same as

$$
{1 \over \kappa_R} = {1 \over N} [{\partial V \over \partial M} + M_0^2
~{\partial ^2V \over \partial M^2}]\mid_{M_0} ~. \eqno(18)
$$

\noindent\bigskip
The solution to this is

$$
{1 \over \kappa} = {1 \over \kappa_R} + {1 \over 4 \pi} \int_{-
\infty}^{\infty} ~{dk_1 \over \omega_0} + \delta X \eqno(19)
$$

\bigskip\noindent
where $\omega_0^2 = k_1^2 + M_0$, and where for our choice of
renormalization prescription, $\delta X = - {1 \over 4 \pi}$. Some other
choice of prescription would yield a different pure number for $\delta X$.

\bigskip
We then have

$$
{1 \over N} {dV \over dM} = {1 \over \kappa_R} - {1 \over 8 \pi} \int_{-
\infty}^{\infty} dk_1 tr \biggr{[}{1 \over \omega }- {1 \over
\omega_0}\biggr{]} + \delta X ~.\eqno(20)
$$

\bigskip\noindent
The gap equation is just the statement that ${dV \over dM}$ vanishes:

$$
\delta X + {1 \over \kappa_R} = {1 \over 8 \pi} \int_{- \infty}^{\infty}
dk_1 tr ~\biggr{[}{1 \over \omega }- {1 \over \omega_0}\biggr{]} ~. \eqno(21)
$$

\bigskip\noindent
Note that $tr ~{1 \over \omega} = {1 \over \sqrt{M + (k_1 + \mu)^2}} + {1
\over \sqrt{M + (k_1 - \mu)^2}}$ which is even in $k_1$. Therefore

\bigskip
$$
\delta X + {1 \over \kappa_R} = {1 \over 4 \pi} J ~~,\quad\quad\quad\quad
{\rm where} ~J~ {\rm is ~the ~integral} \eqno(22)
$$

$$
J = \int_0^{\infty} dk \biggr{[}{ 1 \over \sqrt{M + (k + \mu)^2}} + {1
\over \sqrt{M + (k - \mu)^2}} - {2 \over \sqrt{M_0 + k^2}} \biggr{]} ~.
\eqno(23)
$$

\bigskip\noindent
If we evaluate $J$, we can find the particular $M$ that obeys eqn. (22),
thereby solving the gap equation. We can also obtain the more general
expression

$$
{1 \over N} { dV \over dM} = {1 \over \kappa_R} - {1 \over 4 \pi} J + \delta X
$$

\bigskip\noindent
as a function of $M$, and integrate it to obtain $V(M)$ via eqn. (13).

We find, after some mild computational exertions, that

$$
J(M) = - ln ~M/M_0 \eqno(24)
$$

\noindent
and

$$
{1 \over N} V(M) = ({1 \over \kappa_R} + \delta X)M + {M \over 4 \pi} (ln
M/M_0 - 1) + \vartheta({1 \over N}) ~. \eqno(25)
$$

\bigskip\noindent
As advertised, these expressions are independent of $\mu$. The solution to
the gap equation for our choice of $\delta X$ is

$$
M = M_0e^{(1 - 4 \pi / \kappa_R)} ~. \eqno(26)
$$

\bigskip\noindent
There is no critical lower bound to $\kappa_R$ below which no
solution exists. We see from eqn. (23) that this is due to the fact
that as $M \rightarrow 0$, the expression for $J$ diverges
logarithmically. This infrared singularity is present in $2$
dimensions independent of the value of $\mu$. In higher dimensions,
we expect this singularity to be present at the Fermi surface, $k =
\mid \mu \mid$, and to disappear as $\mu \rightarrow 0$.

We see from eqn. (26) that as $M_0$ is increased for fixed $M$,
$\kappa_R$ becomes smaller. This is an indication that the coupling
$\kappa_R$ is asymptotically free, just as in the original GN
model. In fact, it is not hard to show from the renormalization
condition (19) with cutoff $\Lambda$ that the beta function has the
form

$$
\beta(\kappa) = {-\kappa^2 \over2\pi} ~,
$$

\bigskip\noindent
so that the gap, eq. (26), obeys $(2M_0 {\partial \over \partial
M_0} - \beta(\kappa_R) {\partial \over \partial \kappa_R}) M = 0$.

As a consequence of the Coleman-Mermin-Wagner theorem we expect that terms
which are higher order in ${1 \over N}$ will destabilize the leading order
result, i.e. will give rise to contributions that dominate the ones we have
found for sufficiently large $M$.

We may also [4], [8] compute the Green's functions associated with the
fields $B$ and $B^{\dagger}$. To do this, we need the effective action, not
just the effective potential. This can be read off from equation (10):

$$
\Gamma_{eff} = \int d^4x ({- 4N \over \kappa}) B^{\dagger}(x) B(x) - {i
\over 2} Tr ln [1 + 4A^{-1} B \tilde{A}^{-1} B^{\dagger}] ~. \eqno(27)
$$

Here we have defined $\tilde{A} = \sigma_2 A^T \sigma_2$. Because the gap
equation is independent of $\mu$, we shall simplify our task somewhat by
setting $\mu = 0$; then $A^T = - A$, and

$$
A = i \partial_0 + i \sigma_3 \partial_1 \eqno(28)
$$
$$
\tilde{A} = - i \partial_0 + i \sigma_3 \partial_1 ~. \eqno(29)
$$

\bigskip\noindent
Note that $A \tilde{A} = \tilde{A}A = \partial_0^2 - \partial_1^2$.

To obtain the desired Green's functions, we perform the following sequence
of steps:

\bigskip\noindent
(a) ~We write

$$
B = B_0 + B^{\prime} \eqno(30a)
$$

$$
B^{\dagger} = B_0 + B^{\prime\dagger} ~. \eqno(30b)
$$

\bigskip\noindent
where, as above, $B_0$ is a solution of the gap equation $(M = 4B_0^2)$.

\bigskip\noindent
(b) ~We expand $\Gamma_{eff}$ to second order in $B^{\prime}$ and
$B^{\prime\dagger}$. The linear terms in $B^{\prime}$ and
$B^{\prime\dagger}$ will cancel because of the gap equation. The
coefficients of the quadratic terms will be the inverses of the Green's
functions that we seek.

\bigskip\noindent
(c) ~We observe that by introducing the real and imaginary parts of
$B^{\prime} : ~B^{\prime} = \phi_1 + i \phi_2$, $B^{\prime\dagger} = \phi_1
- i \phi_2$, the off-diagonal terms will disappear; i.e. there will be no
mixed $\phi_1 \phi_2$ terms. In fact, what we find is:

$$
\Gamma_{eff} \simeq - 4N \int d^2x d^2y  \{ \phi_1(x) \phi_1(y) \int {d^2p
\over (2\pi)^2} e^{-ip \cdot (x - y)} [{1 \over \kappa} - {i \over 2}
\Phi_+(p)]
$$
$$
\quad\quad\quad+ \phi_2(x) \phi_2(y) \int {d^2p \over (2\pi)^2} e^{-ip
\cdot (x - y)} [{1 \over \kappa} - {i \over 2} \Phi_-(p)]  \} \eqno(31)
$$

\noindent
where

$$
\Phi_{\pm}(p) = 2 \int {d^2\kappa \over (2 \pi)^2} {[\kappa_0(p_0 +
\kappa_0) - \kappa_1(p_1 + \kappa_1) \pm M] \over [- \kappa_0^2 +
\kappa_1^2 + M] ~[- (\kappa_0 + p_0)^2 + (\kappa_1 + p_1)^2 + M]} ~. \eqno(32)
$$

\bigskip\noindent
Now $\Phi_{\pm}(p)$ is logarithmically divergent, but so is ${1 \over
\kappa}$, and using the gap equation it is easy to see that the divergence
cancels, along with all residual dependence on $\kappa_R$ and $\delta X$.

\bigskip\noindent
(d) ~The integrals defining $\Phi_{\pm}(p)$ can be done explicitly, with
the result that

$$
{\cal G}_{11}(p) \equiv {1 \over \kappa} - {i \over 2} \Phi_+(p) = {1 \over
4\pi\beta} log [{1 + \beta \over 1 - \beta}] \eqno(33)
$$
$$
{\cal G}_{22}(p) \equiv {1 \over \kappa} - {i \over 2} \Phi_-(p) = {\beta
\over 4\pi} log [{1 + \beta \over 1 - \beta}] ~.\eqno(34)
$$

\bigskip\noindent
Here $\beta = \sqrt{{p^2 \over p^2 - 4M}}$, and $p^2 = p_0^2 - p_1^2$.

We see that both ${\cal G}_{11}$ and ${\cal G}_{22}$ give rise to a
branch point at $p^2 = 4M$, and become complex for $p^2 > 4M$,
whereas they are real for $p^2 < 4M$. Furthermore, ${\cal
G}_{22}(p)$ has a simple zero in $p^2$ at $p^2 = 0$, which means
the corresponding Green's function has a pole. Together, these
results suggest that the fermion acquires a mass $m_F = \sqrt{M}$,
while the pole at $p^2 = 0$ is evidence for the existence of a
would-be Goldstone boson that reflects the condensation of
$\left\langle \psi\psi \right\rangle$.

In this note we have analyzed a $2$-dimensional model that exhibits
the formation of Cooper pairs in leading order in ${1 \over N}$.
This condensation occurs for all values of the coupling (as long as
the bare coupling is positive) for any value of the chemical
potential $\mu$, including $\mu = 0$. The coupling itself is
asymptotically free. At $\mu = 0$, we have computed the two-point
functions associated with the composite fields $\psi\psi$ and
$\psi^{\dagger}\psi^{\dagger}$, and have found that $\psi$ acquires
a Majorana mass $2B_0$, where $B_0 = \left\langle \psi\psi
\right\rangle$. We also find evidence for a massless pole, which
indicates the spontaneous breaking of fermion number at large $N$.
However, we expect that the $\langle
\psi\psi \rangle$ condensate will be unstable against  higher order
corrections in ${1 \over N}$, so as not to violate the
Coleman-Mermin-Wagner theorem.  Bearing this in mind, we speculate
that our $GN$-like model at large $N$ could serve as a theoretical
laboratory for a one-dimensional superconductor [9].

\bigskip\noindent
{\bf Acknowledgements}

The research of AC is supported in part by DOE grant DE-FG02-92ER-40704. In
addition,
AC and HM are supported in part by the Grant-in-Aid for International
Scientific Research
 No. 09045036, Inter-University Cooperative Research, Ministry of
Education, Science,
 Sports and Culture of Japan. This work has been performed as an activity
supported by
  the TMU-Yale Agreement on Exchange of Scholars and Collaborations. AC
wishes to acknowledge
   the hospitality of Tokyo Metropolitan University, where part of this
work was begun. FC and HM are grateful for the hospitality of the Center
for Theoretical Physics at Yale.

\bigskip\noindent
{\bf References}

\noindent
1. D. Bailin and A. Love, Phys. Rep. {\bf 107} (1984) 325; M.
Alford, K. Rajagopal and F. Wilczek, Phys. Lett. {\bf 422B} (1998)
247; R. Rapp, T. Sch\"{a}fer, E.V. Shuryak and M. Velkovsky, Phys.
Rev. Lett. {\bf 81} (1998) 53.

\noindent
2. M. Alford, K. Rajagopal and F. Wilczek, hep-ph/9802284 and
hep-ph/9804403; J. Berges and K. Rajagopal, hep-ph/9804233; T.
Sch\"{a}fer, nucl-th/9806064; T. Sch\"{a}fer and F. Wilczek,
hep-ph/9811473.

\noindent
3. N. Evans, S.D.H. Hsu and M. Schwetz, hep-ph/9808444 and
hep-ph/9810514; T. Sch\"{a}fer and F. Wilczek, hep-ph/9810509.

\noindent
4. D.J. Gross and A. Neveu, Phys. Rev. {\bf D10} (1974) 3235.

\noindent
5. S. Coleman, Comm. Math. Phys. {\bf 31} (1973) 259; N.D. Mermin
and H. Wagner, Phys. Rev. Lett. {\bf 17} (1966) 1133.

\noindent
6. E. Witten, Nucl. Phys. {\bf B145} (1978) 110.

\noindent
7. R.L. Stratonovich, Doklady Akad. Nauk. S.S.S.R. {\bf 115} (1957)
1097; S. Coleman, "Aspects of Symmetry," Cambridge Press, 1985, p.
354.

\noindent
8. S. Coleman, R. Jackiw and H.D. Politzer, Phys. Rev. {\bf D10}
(1974) 2491; L. Abbott, J. Kang and H. Schnitzer, Phys. Rev. {\bf
D13} (1976) 2212; C.M. Bender, F. Cooper and G.S. Guralnik, Ann.
Phys. (N.Y.) {\bf 109} (1977) 165.

\noindent
9. For a discussion of Cooper pairing in $1-D$ electron systems,
see for example J. Solyom, Advances in Physics {\bf 28} (1979) 201.

\end